\documentclass[english,aps,prl,twocolumn,superscriptaddress]{revtex4-2}
\usepackage{amssymb}
\usepackage{amsfonts}
\usepackage{subfigure}
\usepackage{graphicx}
\usepackage{dcolumn}
\usepackage{bm}
\usepackage{xr-hyper}
\usepackage{hyperref} 

\usepackage[mathlines]{lineno}
\usepackage{amsmath}
\usepackage{graphicx}
\usepackage{float}
\usepackage{graphicx}
\usepackage{microtype}
\usepackage{overpic} 

\preprint{}
\begin{document}
\title{ Exotic metallic phase in a heterotic Kondo lattice: violation of Luttinger's theorem and absence of Kondo gap}

\author{Hanting Zhong}
\affiliation{School of Physics, Zhejiang University, Hangzhou 310058, China}
\affiliation{School of Physics, Hangzhou Normal University, Hangzhou 310036, China}

\author{Shuxiang Yang}
\affiliation{Zhejiang Lab, Hangzhou 311121, China}

\author{Chao Cao}
\email{ccao@zju.edu.cn}
\affiliation{School of Physics, Zhejiang University, Hangzhou 310058, China}
\affiliation{Center for Correlated Matter, Zhejiang University, Hangzhou 310058, China}
\affiliation{Institute for Advanced Study in Physics, Zhejiang University, Hangzhou 310058, China}

\author{Xiao-Yong Feng}
\affiliation{School of Physics, Hangzhou Normal University, Hangzhou 310036, China}

\author{Jianhui Dai}
\email{daijh@hznu.edu.cn}
\affiliation{School of Physics, Hangzhou Normal University, Hangzhou 310036, China}
\affiliation{Institute for Advanced Study in Physics, Zhejiang University, Hangzhou 310058, China}

\author{Hai-Qing Lin}
\email{hqlin@zju.edu.cn}
\affiliation{School of Physics, Zhejiang University, Hangzhou 310058, China}
\affiliation{Institute for Advanced Study in Physics, Zhejiang University, Hangzhou 310058, China}

\begin{abstract}
  Fermi surface transformation accompanied by quantum phase transition is widely observed in correlated electron systems including heavy fermion materials and van der Waals quantum heterostructures. The unified Anderson lattice description for this phenomenon suggests that their interacting ground states in the case of electron half filling always destabilize the metallic phase in the strong Kondo-coupling regime.  We examine this issue within the paramagnetic phase of a Kondo lattice system with a heterotic layered structure. In the weak Kondo-coupling regime, the system exhibits a fractional Dirac liquid involving the massless or massive Dirac fermions and a charge neutral spinon Fermi surface. In the strong Kondo-coupling regime, the spinons hybridize with the Dirac fermions resulting in an intermediate nearly-flat hybridizing band together with the conduction and valence heavy fermion bands, yielding a large electric Fermi surface. Absence of the Kondo gap between these bands is observed depending on the stacking configurations of the layered structure. In particular, the system remains metallic even at the exact half filling in this regime and Luttinger's theorem is violated across the critical point. Implications of these results on the van der Waals quantum heterostructures are briefly discussed.
\end{abstract}

 \maketitle

\textbf{Introduction.}
Based on the phenomenological reminiscence of flat-band and heavy fermion physics \cite{StewartRMP1984,StewartRMP2001,Colemanbook2007}, a family of quantum heterostructures (QHSs), including the two-dimensional van der Waals and graphene-like twisted bilayers, has been increasingly recognized as generalized Anderson or Kondo lattice systems involving both itinerant and local electronic orbitals \cite{VanoNature2021,ChenPRB2021,RamireaPRL2021,DalalPRR2021,KumarPRB2022,ShiPRB2022,SongPRL2022,ZhaoNature2023,HuSA2023,GuerciSA2023,PoseyNature2024,FumegaNL2024,
ChenNC2024,CheckelskyNRM2024,XiePRR2024,Mahankaliarxiv2025,XiaoNature2026,ZhongPRB2026}. This rekindles the interest in the possible Kondo-effect-driven Fermi surface (FS) transformation \cite{ColemanJPCM2001,SiNature2001,SiScience2010} in these systems. A topological constraint dictating the transformation is the so-called Luttinger theorem, which relates the particle density or Fermi volume enclosed by the FS of metallic many-body systems to the single-particle Green's function evaluated on the interacting ground states \cite{LuttingerPR1960,LuttingerPR1960a, NakajimaPTP1960, MartinPRL1982}. Following the continuity argument of Oshikawa, the $d$-dimensional Fermi volume of a Kondo lattice involving $n_e$ number of electrons and $N_s$ number of local spins of magnitude $S$ per unit cell of volume $v_0$ is given by $V_{\text{FS}} =\frac{(2\pi)^d}{2{v_0}} (n_e + 2N_s S\mod 2)$ \cite{OshikawaPRL2000, OshikawaPRL2000Kondo,HazraPRR2021}. This implies the participation of local spins in the formation of the electric FS in the presence of Kondo coupling. The canonical Kondo lattice with $N_s=1$ and $S=1/2$ then exhibits a Kondo gap without breaking crystalline symmetries, yielding a Kondo insulator at exact half-filling. The argument applies to the FS transformation across a magnetic quantum phase transition in the normal-metal-based Kondo lattices since the unit cell is doubled breaking the translational invariance \cite{VojtaJLTP2010}.

However, the strange metal behaviors observed in QHSs are mostly in proximity to correlation-driven localization-delocalization transitions, suggesting a possible paramagnetic metallic phase as the normal state of interacting electrons at the exact half-filling. Realizing such a metallic ground state necessitates either a violation of Luttinger's theorem as in the non-Fermi liquid due to the single-particle Green's function zeros \cite{DavePRL2013,BlasonPRB2023,GavenskyPRL2023}, or an extension of Oshikawa's argument by including the charge neutral spinon FS volume as in the fractionalized Fermi liquid \cite{SenthilPRL2003,SenthilPRB2004} due to the formation of a quantum spin liquid (QSL)\cite{BalentsNature2010,ZhouRMP2017,KivelsonNRP2023}. The latter implies a paramagnetic Kondo breakdown transition at a finite Kondo coupling $J_c$. This idea has re-emerged in understanding the missing portion of Fermi volume in cuprate superconductors\cite{BonettiRPP2026} with the help of auxiliary local spins described by the ancilla qubit layer model \cite{ZhangPRR2020,ZhangPRB2020} or the CPT model\cite{ColemanPRL2022}. A similar Kondo breakdown transition has been revealed in a 1D spin chain on a semimetallic substrate\cite{DanuPRL2020}.
In this Letter, we study a heterotic Kondo lattice model (HKLM) with three stacking configurations exhibiting the crystalline symmetry pertinent to QHSs. We shall find that this model system unambiguously exhibits a Kondo breakdown transition violating Luttinger's theorem and a paramagnetic metallic state at exactly half-filling in the strong Kondo-coupling regime, thereby providing a platform accommodating the Kondo-effect-driven strange metallicity in QHSs.

\begin{figure*}[tbp]
    \centering
    \includegraphics[width=\textwidth]{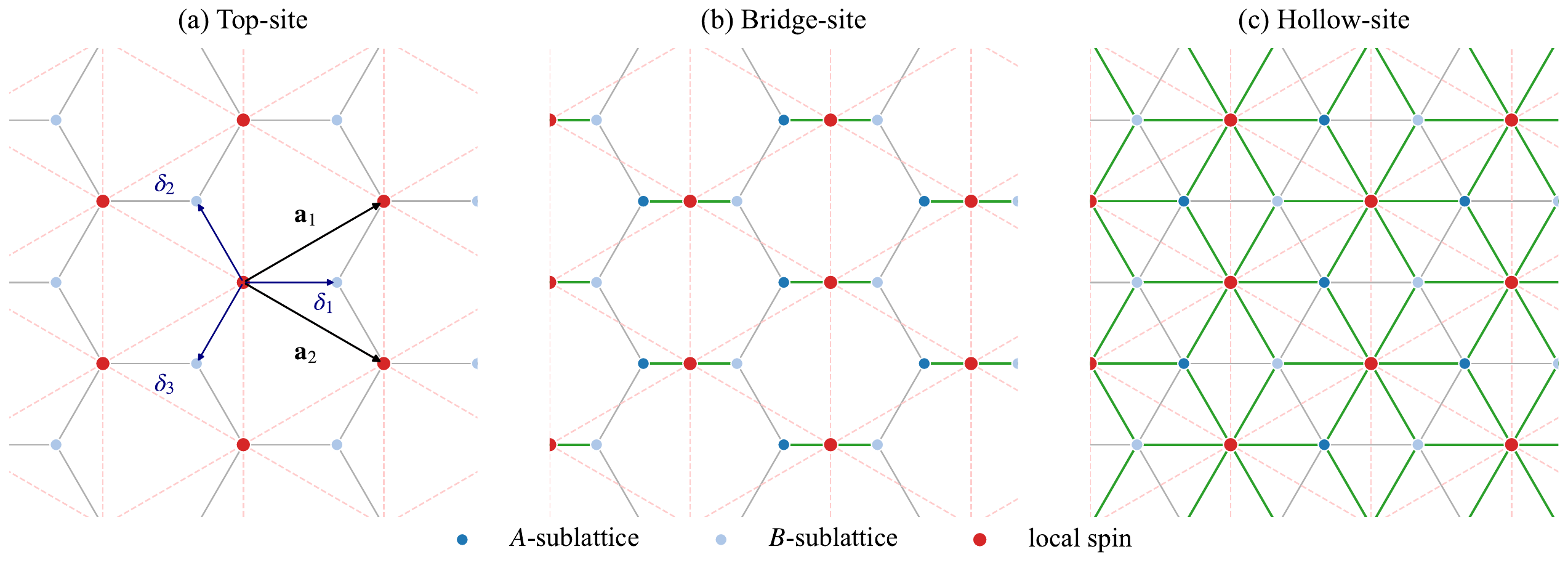}
\caption{Archetypal geometric configurations of the HKLM (top-down view). Light and dark blue circles represent odd ($A$) and even ($B$) sites on the honeycomb lattice, respectively, while red circles denote the sites on the triangular lattice. }
\label{fig:stacking_configs}
\end{figure*}

\textbf{Model Hamiltonian.}
The HKLM we considered consists of three parts: $\hat{\mathcal{H}} = \hat{\mathcal{H}}_c + \hat{\mathcal{H}}_{\text{K}} + \hat{\mathcal{H}}_{\text{H}}$. The first part describes the itinerant electrons moving on the honeycomb lattice:
\begin{eqnarray}
\hat{\mathcal{H}}_c = -t\sum_{\langle\mathbf{i},\mathbf{j}\rangle\alpha}(\hat{c}^{\dagger}_{\mathbf{i}\alpha}\hat{c}_{\mathbf{j}\alpha}+\text{H.c.})-\mu \sum_{\mathbf{i}\alpha}\hat{n}_{c,\mathbf{i}\alpha},
\label{eq:Hc}
\end{eqnarray}
where ${\hat c}^{\dagger}_{\mathbf{i}\alpha}$ creates an electron with spin component $\alpha$ (=$\uparrow$ or $\downarrow$) at the honeycomb lattice site $\mathbf{i}$ in the basal plane ($z=0$),
$\hat{n}_{c,\mathbf{i}\alpha} = \hat{c}^{\dagger}_{\mathbf{i}\alpha}\hat{c}_{\mathbf{i}\alpha}$ is the number operator, $t$  the nearest-neighbor in-plane hopping amplitude, and $\mu$ the chemical potential.
The second part describes the interlayer nearest-neighbor Kondo coupling:
\begin{eqnarray}
\hat{\mathcal{H}}_{\text{K}} = J_K \sum_{\langle \mathbf{i}, \mathbf{I} \rangle} \hat{\mathbf{s}}_{c,\mathbf{i}} \cdot \hat{\mathbf{S}}_{\mathbf{I}}~,
\end{eqnarray}
where $\hat{\mathbf{s}}_{c,\mathbf{i}} = \frac{1}{2} \sum_{\alpha, \beta} \hat{c}^{\dagger}_{\mathbf{i}\alpha} \boldsymbol{\sigma}_{\alpha\beta} \hat{c}_{\mathbf{i}\beta}$ with $\boldsymbol{\sigma}=(\sigma_x,\sigma_y,\sigma_z)$ being the Pauli matrices, $\hat{\mathbf{S}}_{\mathbf{I}}$ is local spin of $1/2$ at the triangular lattice site $\mathbf{I}$ in the plane $z=-a_z$.
The third part describes the Heisenberg-like interactions of local spins:
\begin{eqnarray}
\hat{\mathcal{H}}_{\text{H}} = J_H \sum_{\langle\mathbf{I}, \mathbf{J} \rangle} \hat{\mathbf{S}}_{\mathbf{I}} \cdot \hat{\mathbf{S}}_{\mathbf{J}}+ J'_H\sum_{\langle\langle\mathbf{I}, \mathbf{J} \rangle\rangle} \hat{\mathbf{S}}_{\mathbf{I}} \cdot \hat{\mathbf{S}}_{\mathbf{J}}.
\label{eq:RKKY}
\end{eqnarray}
This part could be dynamically generated via the interlayer coupling and the antiferromagnetic $J_H$ and $J'_H$ represent the nearest-neighbor and the next-nearest-neighbor effective Ruderman-Kittel-Kasuya-Yosida (RKKY) interactions, respectively.

The present HKLM can be mapped onto a kagomelike system involving itinerant honeycomb and the local moment triangular lattices assigned in the $z=0$ and $z=-a_z$ planes, respectively, sharing the same 2D Bravais structure with the planar lattice vectors $\mathbf{a}_1 = a_0(3/2, \sqrt{3}/2)$ and $\mathbf{a}_2 = a_0(3/2, -\sqrt{3}/2)$ ($a_0$ is the distance between nearest neighboring sites). Each site on the honeycomb lattice is labelled by $\mathbf i=(\mathbf m, \eta)$, with $\mathbf m=(m_1,m_2)$ being a pair of integers specifying the unit cell and $\eta=A,B$ specifying the odd or even honeycomb sublattices, respectively. Its position is $\mathbf{r}_{\mathbf{i}} =(m_1\mathbf{a}_1 + m_2\mathbf{a}_2 + \mathbf r_\eta, 0)$, with $\mathbf r_A = (0,0)$ and $\mathbf r_B = a_0 (1,0)$.
Similarly, each site on the triangular lattice is labelled by $\mathbf I=(\mathbf m,f)$ with the position $\mathbf{R}_{\mathbf{I}} = (m_1\mathbf{a}_1 +m_2 \mathbf{a}_2 + \mathbf r_f, -a_z)$ and $a_z$ the interlayer distance. Here the vector $\mathbf r_f$ determines the stacking configurations as depicted in Fig. \ref{fig:stacking_configs}: (a) the top-site (\textbf{T}) configuration, $\mathbf r_f=0$ with the nearest-neighbor Kondo bonds ( the interlayer coordination number) $Z_{\mathbf I}=Z=1$ and  $C_{3v}$ symmetry; (b) the bridge-site (\textbf{B}) configuration, $\mathbf r_f=\frac{a_0}{2}(1,0)$ with $Z=2$ and $C_{2v}$ or ${\cal M}_y$ symmetry; (c) the hollow-site (\textbf{H}) configuration, $\mathbf r_f=-a_0(1,0)$ with $Z=6$ and $C_{6v}$ symmetry.

\begin{figure*}[t]
    \centering
    \includegraphics[width=0.33\textwidth]{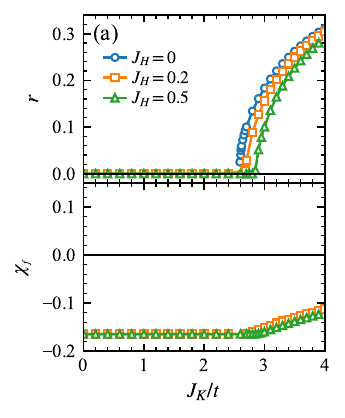}\hfill
    \includegraphics[width=0.33\textwidth]{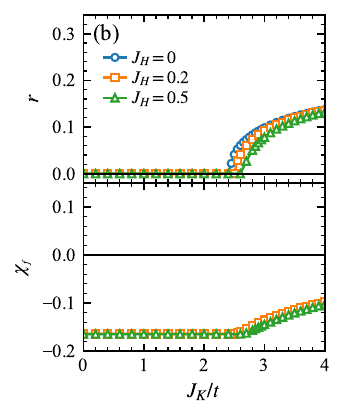}\hfill
    \includegraphics[width=0.33\textwidth]{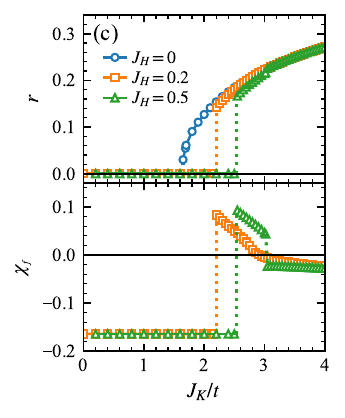}
    \caption{Kondo and spinon order parameters. The Kondo hybridization $r$ (upper panels) and the effective spinon hopping $\chi_f$ (lower panels) for (a) \textbf{T}-configuration, (b) \textbf{H}-configuration, and (c) \textbf{B}-configuration, respectively.}
    \label{fig:Order_Params}
\end{figure*}

\textbf{Transition from weak to strong Kondo-coupling regimes.} In the weak-coupling limit $J_K=0$, the low energy states of the itinerant electrons at the half-filling ($\mu=0$) are free Dirac fermions exhibiting the linear dispersion around the Dirac points. The local spins are expected in a QSL  state when $J'_H/J_H= 0.07 \sim 0.17$ \cite{ManulPRB1999,ZhuPRB2015,IqbalPRB2016,HuPRL2019}. Given the small $J'_H$ required for this putative state we shall neglect $J'_H$ without changes in main conclusions and leave a discussion on its numerical influence in Supplemental Material (SM)\cite{SM}.
 The local spins are then represented by the Abrikosov pseudo-fermions \cite{AbrikosovPRF1965}, $\hat{\mathbf{S}}_{\mathbf{I}} = \frac{1}{2}\sum_{\alpha\beta}\hat{f}_{\mathbf{I}\alpha}^{\dagger}\boldsymbol{\sigma}_{\alpha\beta}\hat{f}_{\mathbf{I}\beta}$, subject to the single-occupancy constraint $\sum_{\alpha}\hat{f}_{\mathbf{I}\alpha}^{\dagger}\hat{f}_{\mathbf{I}\alpha}= 1$. The constraint leads to a local $U(1)$ gauge symmetry under $\hat f_{\mathbf{I}} \rightarrow \hat f_{\mathbf{I}} e^{i\theta_{\mathbf{I}}}$. Physically, $\hat f_{\mathbf{I}}$ describes the neutral fermionic spinons, the excitations carrying spin-1/2 on the posited ground state $|G\rangle$\cite{BalentsNature2010,LeeRMP2006}. Mathematically, the constraint is enforced by introducing a Lagrange multiplier $\lambda_f(\mathbf I)(\sum_{\alpha}\hat{f}_{\mathbf{I}\alpha}^{\dagger}\hat{f}_{\mathbf{I}\alpha}-1)$ in the Hamiltonian. The energy gain from the RKKY interaction could break the gauge symmetry in low energy excitations characterized by a nonzero expectation value $\chi_{f}(\mathbf I \mathbf J)=\frac{1}{2}\langle G| \hat f^{\dagger}_{\mathbf{I}}\hat f_{\mathbf{J}}|G\rangle$. Hence the Dirac semimetal and the $U(1)$ QSL coexist in the limit $J_K=0$.

 In the strong-coupling limit $J_K\rightarrow \infty$, the itinerant electrons and spinons combine into the bosonic Kondo pairs $\hat V_{\mathbf i \mathbf I} \equiv \sum_{\alpha}\hat f^{\dagger}_{\mathbf{I}\alpha}\hat c_{\mathbf{i}\alpha}$ such that
\begin{eqnarray}
 \hat{\mathcal{H}}_{\text{K}} = \sum_{\langle \mathbf{i}, \mathbf{I} \rangle}[ -\frac{J_K}{2}
    \hat V^{\dagger}_{\mathbf i \mathbf I} \hat V_{\mathbf i \mathbf I}
 + \frac{J_K}{4}Z_{\mathbf I} \sum_{\alpha}\hat{n}_{c,\mathbf{i}\alpha}].
\label{eq:H_Kondo_Fierz}
\end{eqnarray}
 The first term favours the condensation of the Kondo pairs, i.e.,  $r(\mathbf i \mathbf I)=\frac{1}{2}\langle G|\hat V_{\mathbf i \mathbf I}|G\rangle =\frac{1}{2}|V_{\mathbf i \mathbf I}| e^{i\theta_{\mathbf i \mathbf I}}\neq 0$, leading to spontaneous breaking of the gauge symmetry via the so-called Anderson-Higgs mechanism \cite{FradkinPRD1979,LeeRMP2006,Colemanbook2016,RaczkowskiPRB2022}. The second term acts as a configuration-dependent local potential for the itinerant electrons.
Importantly, the Kondo condensate dynamically locks the temporal phase of the Kondo pairs and the spinons participate in the global $U(1)$ charge conservation. A large electric FS contributed by both itinerant electrons and spinons is then expected.

When $J_K$ is finite, the above two limiting tendencies compete and a critical point $J_c$ signalling the Kondo condensate is expected, separating the weak ($J_K<J_c$) and strong ($J_K>J_c$) coupling regimes. Remarkably, $J_c$ is not accessible from the $J_K=0$ limit using the adiabatic continuity argument because the Kondo coupling is irrelevant in either Dirac liquid or $U(1)$ QSL\cite{note2}, yet the local potential could open a gap in the Dirac liquid \cite{SemenoffPRL1984,NetoRMP2009} depending on the symmetry of stacking configurations.

Here, we solve $J_c$ by the mean-field method assuming the uniform expectation values $\chi_{f}(\mathbf I \mathbf J)$ and $r(\mathbf i \mathbf I)$ for nearest neighboring pairs, represented by two bosonic order parameters $\chi_f$ and $r$, respectively, together with the uniform Lagrange multiplier $\lambda_f$. The Kondo and RKKY interactions are then decoupled in the particle-hole channel, obtaining the following effective mean-field Hamiltonian matrix in the momentum space:
\begin{equation}
\resizebox{0.907\columnwidth}{!}{$
\displaystyle
H_{\mathbf{k}} =
\begin{pmatrix}
        -\mu + \Delta E_A & -t \gamma_{\mathbf{k}} & -J_K r \mathcal{W}_A(\mathbf{k}) \\
        -t \gamma_{\mathbf{k}}^* & -\mu + \Delta E_B & -J_K r \mathcal{W}_B(\mathbf{k}) \\
        -J_K r \mathcal{W}_A^*(\mathbf{k}) & -J_K r \mathcal{W}_B^*(\mathbf{k}) & \lambda_f + \varepsilon_{f,\mathbf{k}}
\end{pmatrix}.
$}
\label{eq:H_3x3}
\end{equation}
In the above, $\gamma_{\mathbf{k}} = \sum_{j=1}^3 e^{i\mathbf{k}\cdot\boldsymbol{\delta}_j}$ with $\boldsymbol{\delta}_1 = a_0(1, 0)$, $\boldsymbol{\delta}_2 = a_0(-1/2, \sqrt{3}/2)$, $\boldsymbol{\delta}_3 = a_0(-1/2, -\sqrt{3}/2)$,
and $\varepsilon_{f,\mathbf{k}} = -2J_H \chi_f F_{\mathbf{k}}$ with $F_{\mathbf{k}} = \cos(\mathbf{k}\cdot\mathbf{a}_1) + \cos(\mathbf{k}\cdot\mathbf{a}_2) + \cos[\mathbf{k}\cdot(\mathbf{a}_1-\mathbf{a}_2)]$. The local potential $\Delta E_\eta$ and form factor $\mathcal{W}_\eta(\mathbf{k})$ are defined as: \textbf{T}-configuration, $\Delta E_A = J_K/4$, $\Delta E_B = 0$, $\mathcal{W}_A = 1$, $\mathcal{W}_B = 0$; \textbf{B}-configuration, $\Delta E_A = \Delta E_B = J_K/4$, $\mathcal{W}_A = e^{i k_x a_0 / 2}$, $\mathcal{W}_B = e^{-i k_x a_0 / 2}$; \textbf{H}-configuration, $\Delta E_A = \Delta E_B = 3J_K/4$, $\mathcal{W}_A = \gamma_{\mathbf{k}}^*$, $\mathcal{W}_B = \gamma_{\mathbf{k}}$.
The three quasiparticle bands are solved as the eigenvalues $E_{\mathbf k \sigma m}$ ($m=1,2,3$) of $H_{\mathbf{k}}$ as detailed in SM \cite{SM}.
 The free energy at temperature $T$ is obtained by  ${\mathcal F}=-\frac{1}{\beta} \sum_{\mathbf k \sigma m}\ln \left[ 1+ e^{-\beta E_{\mathbf k \sigma m}}\right] +E_C$, with $E_C = -L\lambda_f + 2 Z L J_K r^2 + 6LJ_H \chi_f^2 + \frac{3}{4}LJ_H$, $\beta =1/(k_B T)$, $L$ the total number of unit cells. Meanwhile, $\mu$ and $\lambda_f$ are determined by $-\sum_{\mathbf{k}\sigma m} n_F(E_{\mathbf{k}\sigma m})\frac{\partial E_{\mathbf{k}\sigma m}}{\partial \mu}= N$ and
$\sum_{\mathbf{k}\sigma m} n_F(E_{\mathbf{k}\sigma m}) \frac{\partial E_{\mathbf{k}\sigma m}}{\partial \lambda_f} = L$, respectively, with $n_F(E) = (e^{\beta E} + 1)^{-1}$. Minimizing $\mathcal{F}$ with respect to $(r, \chi_f)$ yields a pair of self-consistent mean-field equations:
\begin{eqnarray}
\sum_{\mathbf{k}\sigma m} n_F(E_{\mathbf{k}\sigma m}) \frac{\partial E_{\mathbf{k}\sigma m}}{\partial r} + 4 Z L J_K r &= 0, \label{eq:sc_r} \\
    \sum_{\mathbf{k}\sigma m} n_F(E_{\mathbf{k}\sigma m}) \frac{\partial E_{\mathbf{k}\sigma m}}{\partial \chi_f} + 12L J_H \chi_f &= 0. \label{eq:sc_chi}
\end{eqnarray}
The existence of a critical $J_c$ is then determined by the solution of the following equation
\begin{eqnarray}
J_c=-\frac{1}{4ZL}\lim_{r\rightarrow 0^{+}} \lim_{T\rightarrow 0}\sum_{\mathbf{k}\sigma m} n_F(E_{\mathbf{k}\sigma m})\frac{1}{r} \frac{\partial E_{\mathbf{k}\sigma m}}{\partial r}.
\end{eqnarray}
In the continuous limit ($L\rightarrow \infty$) and at the half-filling ($N=2L$) for $J_H=0$, we obtain $J_c/t \approx 2.564, 2.494$, and $1.674$ for the \textbf{T}-, \textbf{H}-, and \textbf{B}-configurations, respectively, as illustrated in SM \cite{SM}. These results are in agreement with the single-impurity Kondo problem in a pseudo-gap bath\cite{WithoffPRL1990,GonzalezPRB1998}. The value of $J_c$ increases slightly with $J_H$ and decreases prominently with doping as shown below in the numerical solutions.

\textbf{Numerical Results.}
The mean-field equations are numerically solved on a dense momentum mesh at a sufficiently large inverse temperature ($\beta = 400$) approaching the zero-temperature limit ( $t=1$ and $a_0=1$ for convenience). The main results are summarized as follows.

(1)\textit{Kondo and spinon order parameters.}
$r$ and $\chi_f$  with varying $J_K$ are plotted in Fig. \ref{fig:Order_Params} for several values of $J_H$.
For \textbf{T}- and \textbf{H}-configurations, they evolve smoothly from the decoupled phase ($r=0$) into the Kondo phase ($r>0$), unambiguously signifying a continuous transition at $J_c$ estimated previously. For the \textbf{B}-configuration, however, while $r$ is continuous for $J_H=0$, it shows a jump at $J_c$ for $J_H\neq 0$, featuring a weakly first-order transition triggered by the concomitant jump in $\chi_f$ as shown in Fig. \ref{fig:Order_Params} (c). This situation represents a lattice extension of the Kondo impurity problem involving two symmetric channels where the overscreened impurity induces the non-Fermi liquid behavior\cite{NozieresJP1980,AndreiPRL1984,TsvelickJSP1985}.

\begin{figure}[htbp]
    \centering
    \includegraphics[width=\columnwidth]{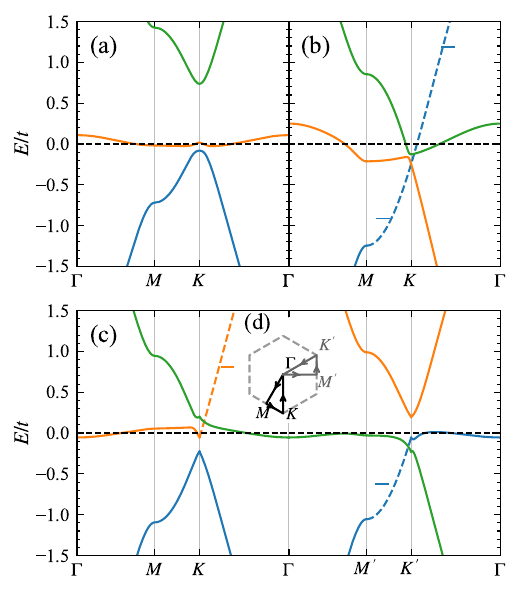}
    \caption{Band structures in the strong-coupling regime. The solved mean-field parameters $(\mu,\lambda_f,r,\chi_f)$ are $(-0.0166, 0.0123, 0.0694,-0.1610)$ for (a) \textbf{T}-configurations ($J_K=2.7$); $(2.3448, -0.0840, 0.0784, -0.1431)$ for (b) \textbf{H}-configuration( $J_K=2.8$); and $(0.5176,-0.0059, 0.0796, 0.1051)$ for (c) \textbf{B}-configuration ($J_K=1.85$), respectively, all with a fixed $J_H=0.1$. (d) The Brillouin zone highlighting the inequivalent momentum paths (orange and blue) associated with the \textbf{B}-configuration, while a single momentum path ($\Gamma$-$M$-$K$-$\Gamma$) is sufficient in (a) and (b) due to the $C_3$ symmetry.}
    \label{fig:Bands}
\end{figure}

(2)\textit{Band structures.}
As shown in Fig. \ref{fig:Bands}, the system inherently remains metallic even at half-filling, manifesting a violation of Luttinger's theorem. Specifically, there is an intermediate hybridizing band (orange) close to the Fermi level in the \textbf{T}-configuration. This band is nearly-flat and gapped from the other two hybridizing (upper/lower) bands in the presence of the staggered local potential due to the lack of the inversion symmetry\cite{ZhongPRB2026}.
In the \textbf{H}-configuration, the $C_{6v}$ symmetry protects the Dirac touching at $K$ without a Kondo gap. A band crossing appears in the $K$-$\Gamma$ direction along which ${\cal W}_{A,B}(\mathbf k)$ vanish, resulting in the decoupled antibonding band ($-$) \cite{SM}. The \textbf{B}-configuration breaks the $C_{3v}$ symmetry lifting the degeneracy at the Dirac points. Meanwhile, the mirror symmetry ${\cal M}_y$ enables the decoupled odd-parity band ($-$) to intersect one of the other hybridizing bands along the lines $K$-$\Gamma$ and $M'$-$K'$, respectively\cite{SM}.

\begin{figure}[t]
    \centering
    \includegraphics[width=0.48\columnwidth]{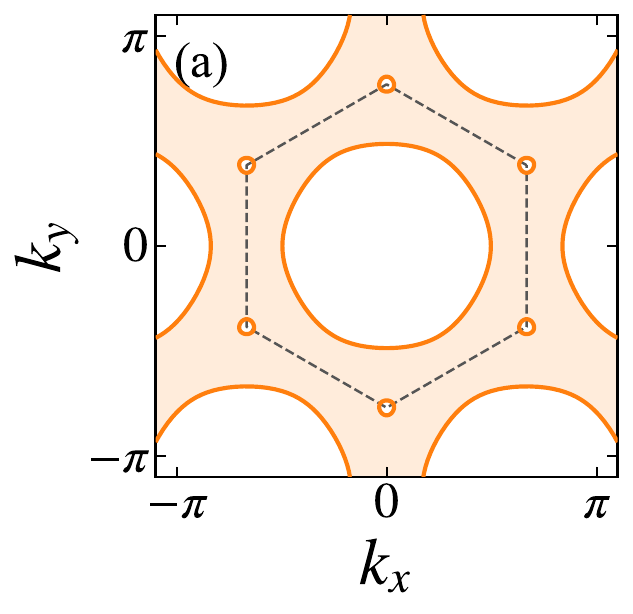} \hfill
    \includegraphics[width=0.48\columnwidth]{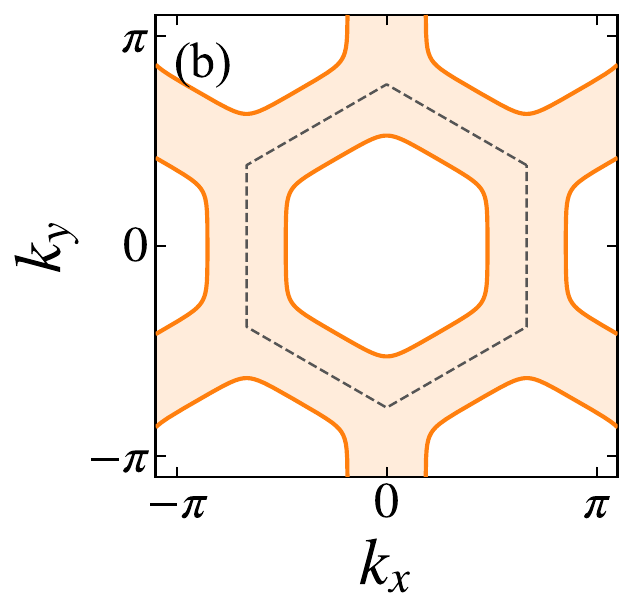} \\ \vspace{0.15cm}
    \includegraphics[width=0.48\columnwidth]{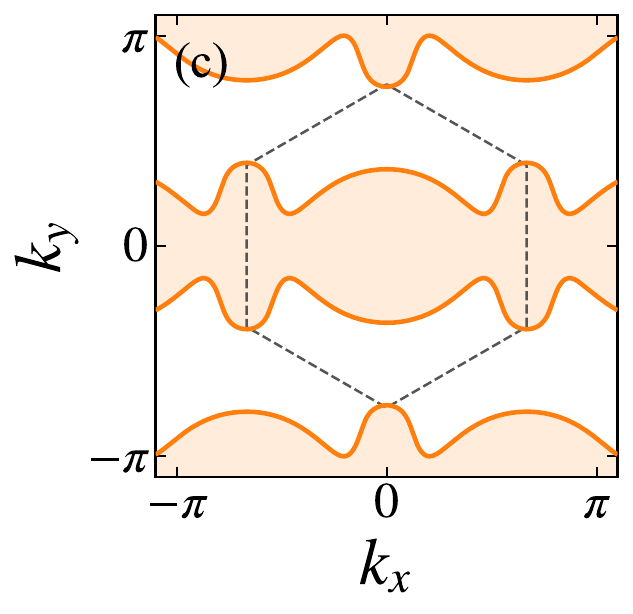} \hfill
    \includegraphics[width=0.48\columnwidth]{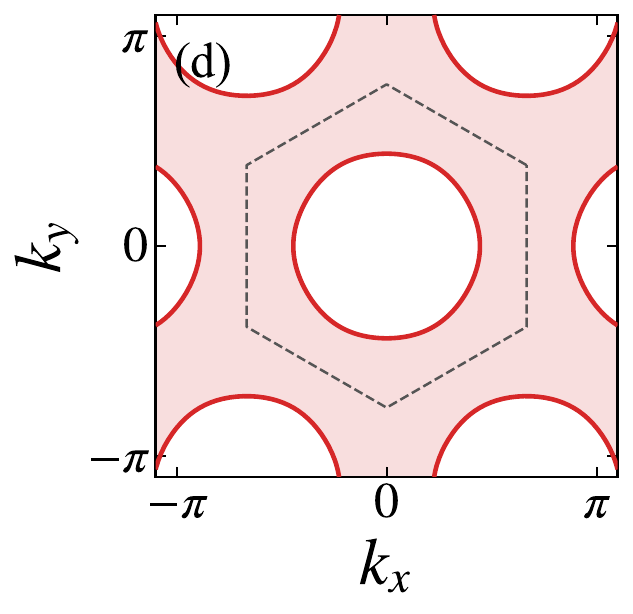}
    \caption{
    Electric and spinon Fermi surfaces. The Fermi surfaces (thick lines ) in the strong-coupling regime for \textbf{T}-,\textbf{H}-, and \textbf{B}-configurations shown in (a),(b), and (c), respectively. The shaded/blank areas denote the occupied/empty states in the intermediate heavy fermion band. The dashed line indicates the first Brillouin zone.  (d) The spinon Fermi surface in the weak-coupling regime. }
    \label{fig:FermiSurfaces}
\end{figure}

(3)\textit{Fermi surfaces.}
In the weak-coupling regime, the itinerant electrons remain in the Dirac liquid at half-filling without a finite FS, indicative of a vanishing electric Fermi volume. Figs. \ref{fig:FermiSurfaces}(a,b,c) display the electric FSs (thick lines) in the strong-coupling regime with the same parameters in the band structures. The shaded/blank areas denote the occupied/empty states of the intermediate heavy band at half-filling. There are small circled blank areas around $K$ in \ref{fig:FermiSurfaces}(a), and the FS in \ref{fig:FermiSurfaces}(c) exhibits the nematic topology due to the mirror symmetry\cite{Fradkin2010}. These results are in accordance with a large electric Fermi volume $V_{\text{FS}}=\frac{(2\pi)^2}{
2v_0}(n_e +2 N_s S\mod 2)=\frac{(2\pi)^2}{2v_0}$ for $S=1/2$ and $n_e=2, N_s=1 $ per unit cell at half-filling. The jump in the Fermi volume from the weak- to strong-coupling regimes indicates the violation of Luttinger's theorem across the transition.
As a comparison, the neutral spinon FS in the weak-coupling regime is plotted in Fig. \ref{fig:FermiSurfaces}(d). It provides the missing portion of Fermi volume in this regime due to the formation of a QSL \cite{BonettiRPP2026}.

\begin{figure}[htbp]
    \centering
    \includegraphics[width=\linewidth]{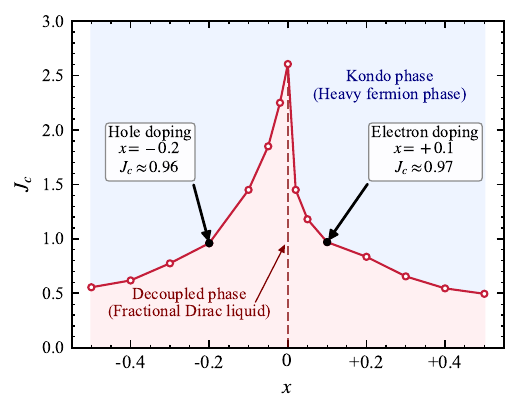}
    \caption{Phase diagram. Critical Kondo coupling $J_c$ vs doping concentration $x$ for the \textbf{T}-configuration. }
    \label{fig:Phase_Diagram}
\end{figure}

(4)\textit{Phase diagram.}
  We have also calculated the critical Kondo coupling as a function of carrier doping concentration $x$ away from half-filling for the \textbf{T}-configuration as shown in Fig. \ref{fig:Phase_Diagram}. The calculated $J_c(x)$ (denoted by small open circle) exhibits a peak at half-filling ($x=0$) and decreases significantly with increasing $|x|$, manifesting the tunable accessibility of the Kondo phase. The phase boundary reveals a striking electron-hole asymmetry as marked by the small black points.

\textbf{Summary and Discussion. }

We have demonstrated a Kondo-effect-driven transition in the kagomelike Kondo lattices with a heterotic structure, providing a conceptually transparent mechanism for a robust breakdown of Luttinger's theorem in the correlated Dirac metals. The pristine lattice structure reveals a formal topological loophole that exempts the fractionalized spinons from Oshikawa's adiabatic continuity argument in the weak-coupling regime, thereby stabilizing a fractional Dirac liquid. In the strong-coupling regime, the system shows an exotic metallic phase at half-filling with a large electric FS due to the formation of an intermediate hybridizing band which is separated from or connected to the other two hybridizing bands with or without a Kondo gap depending on the symmetry of stacking configurations.
Future investigations on the critical fluctuations and ordering instabilities in this exotic metallic phase are desirable.

The present HKLM aligns with a surge of recent experimental advances unifying flat bands, strange metals, and the Kondo effect \cite{CheckelskyNRM2024}. With the kagomelike lattice hosting both the electric Dirac liquid and the neutral QSL, the mechanism revealed here points to a possible universal origin of quantum critical strange metals with a jump in the Fermi volume \cite{PhillipsScience2022, HuNP2024}. This is evidenced in the recent spectroscopic observations of quasiparticle interference suppression \cite{SouzaNP2026} as well as the multipartite entanglement witnessed by quantum Fisher information during Kondo destruction \cite{MazzaNP2026}.  The lattice geometry provides a microscopic analogue to the emergent localized moments in paramagnetic $d$-orbital kagome metals \cite{YeNP2024,SouzaNP2026}. The configuration-dependent band structures and FSs shed light on understanding the direction-dependent Fermiology discovered in bulk aperiodic moir\'e metals \cite{NuckollsNature2026}. Finally, we hope that the HKLM could be realized in the graphene/$\text{WSe}_2$ heterostructure where the strong interlayer hybridization has been evidenced by the scanning tunneling microscopy tip manipulation\cite{ZhangPRL2026}.

We would like to thank Chenchao Xu, Yang Liu, and Huiqiu Yuan for helpful discussions. This work was supported in part by the National Science Foundation of China ( No.12274109 and No. 12274364).


\clearpage
\onecolumngrid 

\begin{center}
    \textbf{\large Supplementary Material for ``Exotic metallic phase in a heterotic Kondo lattice: violation of Luttinger's theorem and absence of Kondo gap''}\\[0.3cm]
    Hanting Zhong, Shuxiang Yang, Chao Cao$^*$, Xiao-Yong Feng, Jianhui Dai$^\dagger$, and Hai-Qing Lin$^\ddagger$
\end{center}
\vspace{0.6cm}

\setcounter{equation}{0}
\setcounter{figure}{0}
\setcounter{table}{0}
\setcounter{page}{1}
\renewcommand{\theequation}{S\arabic{equation}}
\renewcommand{\thefigure}{S\arabic{figure}}
\renewcommand{\thetable}{S\arabic{table}}
\renewcommand{\thepage}{S\arabic{page}} 

\section*{SM-A: Influences of the next-nearest-neighbor spin interaction $J_H^\prime$}

As introduced in Eq.~(\ref{eq:RKKY}) of the main text, the Heisenberg-like RKKY interactions of the local spins include both the nearest-neighbor (NN) term $J_H$ and the next-nearest-neighbor (NNN) term $J_H^\prime$. While a finite $J_H^\prime$ ($J_H^\prime / J_H \approx 0.07 \sim 0.17$) is required to stabilize the putative $U(1)$ quantum spin liquid (QSL) state on the triangular lattice \cite{SMManulPRB1999,SMZhuPRB2015,SMIqbalPRB2016,SMHuPRL2019}, $J_H^\prime = 0$ is used for simplicity in the main text. Here, we consider the numerical influences of nonzero $J_H^\prime$ on $r$ and $\chi_f$.

Following the Abrikosov pseudo-fermion mean-field decoupling utilized in the main text, the NNN spin interaction generates an additional next-nearest-neighbor spinon order parameter:
\begin{eqnarray}
    \chi_f^\prime = \frac{1}{2}\langle \hat{f}_{\mathbf{I}}^\dagger \hat{f}_{\mathbf{J}} \rangle_{\langle\langle \mathbf{I},\mathbf{J} \rangle\rangle}.
\end{eqnarray}
This results in an extra dispersion term $\varepsilon_f^\prime(\mathbf{k}) = -2J_H^\prime \chi_f^\prime F_{\mathbf{k}}^\prime$ for the pseudo $f$-fermions ( or the $f$-spinons in the QSL phase), with
\begin{eqnarray}
F_{\mathbf{k}}^\prime = \cos(3k_x) + 2\cos(\frac{3}{2}k_x)\cos(\frac{3\sqrt{3}}{2}k_y).
\end{eqnarray}
Minimizing the free energy with respect to ($r,\chi_f, \chi_f'$) leads to the modified mean-field equations. We solved these equations for the T-configuration in the range $J_H^\prime / J_H \in [0, 0.3]$ with $J_H = 0.5$.

\begin{figure*}[htbp]
    \centering
    \includegraphics[width=0.9\textwidth]{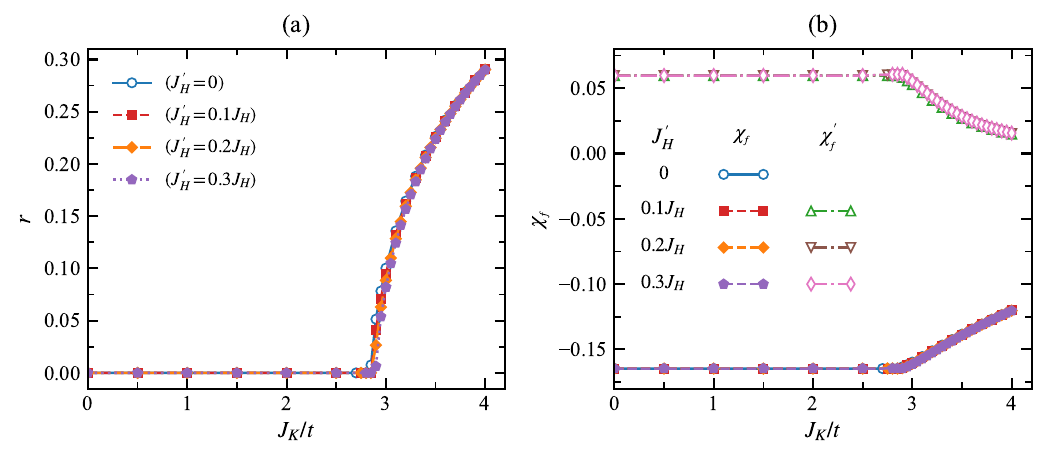}
    \caption{\textbf{Influences of next-nearest-neighbor interactions.} Numerical evaluation of the mean-field order parameters for the \textbf{T}-configuration at a fixed nearest-neighbor interaction $J_H = 0.5$, systematically comparing cases with $J_H^\prime/J_H = 0, 0.1, 0.2$, and $0.3$. (a) Evolution of the Kondo order parameter $r$ as a function of the Kondo coupling $J_K/t$. The curves perfectly overlap, demonstrating the robustness of the Kondo phase. (b) The nearest-neighbor ($\chi_f$) and next-nearest-neighbor ($\chi_f^\prime$) spinon order parameters versus $J_K/t$.}
    \label{fig:JH_prime}
\end{figure*}

As shown in Fig.~\ref{fig:JH_prime}(a), the Kondo order parameter $r$ is almost unchanged upon the introduction of $J_H^\prime$ and the critical Kondo coupling $J_c$ experiences a very slight increase as $J_H^\prime$ grows in the range $J_H^\prime / J_H \in [0, 0.3]$. This shows that the NNN RKKY interaction has very little influence on the formation of the Kondo state although it plays a role in the formation of a QSL in the decoupled phase.

Furthermore, Fig.~\ref{fig:JH_prime}(b) demonstrates that the NN spinon hopping $\chi_f$ is largely unaffected by $J_H^\prime$ across the entire range of $J_H^\prime/J_H$. Although a finite NNN spinon $\chi_f^\prime$ is generated, its magnitude is relatively small compared to $\chi_f$ and is further suppressed by $J_K$ in the Kondo phase. Quantitatively, the NNN RKKY contribution to the spinon dispersion is bounded by $J_H^\prime |\chi_f^\prime| \approx 0.009 t$ for the largest simulated value $J_H^\prime = 0.3 J_H$.  This is an order of magnitude smaller than the contribution from the NN RKKY interaction $J_H |\chi_f| \approx 0.082 t$. This again suggests that the energetic contribution of the NNN RKKY interaction is a minor perturbation. Therefore, $J_H^\prime$ can be safely neglected as long as the Fermi surface transformation from the fractional Dirac liquid to the Kondo phase is concerned.

\section*{SM-B: Eigenvalues of the Hamiltonian matrix}

The three quasiparticle bands are solved as the eigenvalues $E_{\mathbf{k}\sigma m}$ ($m=1,2,3$) of the spin-degenerate mean-field Hamiltonian matrix $H_{\mathbf{k}}$ presented in Eq.~(\ref{eq:H_3x3}) of the main text. The zeros of the secular determinant $\det(E I_{3\times3} - H_{\mathbf{k}})$ lead to the standard cubic equation for the energy eigenvalues $E$:
\begin{equation}
    E^3 + b E^2 + c E + d = 0,
\end{equation}
where the momentum-dependency of $E$ is implied in the representation-independent coefficients:
\begin{align}
    b = -\mathrm{Tr}(H_{\mathbf{k}}), ~~~
    c = \frac{1}{2}\left[ (\mathrm{Tr}H_{\mathbf{k}})^2 - \mathrm{Tr}(H_{\mathbf{k}}^2) \right], ~~~
    d = -\det(H_{\mathbf{k}}).
\end{align}

Following the trigonometric solutions of the cubic equation \cite{SMCardano2007, SMLambert1906}, the hybridized quasiparticle bands are given by
\begin{align}
    E_{\mathbf{k}\sigma m} &= -\frac{b}{3} + 2\sqrt[3]{\rho} \cos \left[\theta -(m-1)\varphi \right],
\end{align}
with the auxiliary intermediate parameters
\begin{equation}
    \varphi=\frac{2\pi}{3},\quad \rho = \sqrt{-\left(\frac{p}{3}\right)^3}, \quad \theta = \frac{1}{3} \arccos\left(-\frac{q}{2\rho}\right),
\end{equation}
and
\begin{equation}
    p = \frac{3c - b^2}{3}, \quad q = \frac{27d - 9bc + 2b^3}{27}.
\end{equation}

The band index $m$ corresponds to the upper ($m=1$), the intermediate($m=2$), and the lower ($m=3$) bands in the manifold of the calculated model parameters. However, band crossings could take place at some $\mathbf {k}$-points due to the configuration-dependent symmetry.

\section*{SM-C: Evaluation of the critical Kondo coupling $J_c$}

\subsection*{1. \textbf{T}-configuration}

While the existence of a critical $J_c$ is predicted for the transition from the decoupled fractional Dirac liquid to the Kondo screened heavy fermion phase, its precise location within the mean-field theory is determined by the solution of the following equation
\begin{eqnarray}
J_c=-\frac{1}{4ZL}\lim_{r\rightarrow 0^{+}} \lim_{T\rightarrow 0}\sum_{\mathbf{k}\sigma m} n_F(E_{\mathbf{k}\sigma m})\frac{1}{r} \frac{\partial E_{\mathbf{k}\sigma m}}{\partial r}.
\end{eqnarray}
Our goal is then to analytically solve $J_c$ in the continuous limit ($L\rightarrow \infty$) and at the half-filling ($N=2L$) for $J_H=0$.

Notice that in the decoupled phase $J_K<J_c$ or $r=0$ in the \textbf{T}-configuration,  the itinerant electrons suffer the staggered potential $J_K/4$ exclusively on the A-sublattice, yielding the unperturbed conduction bands $E_{c\pm}^{(0)} = -\mu + \frac{J_K}{8}\pm \zeta_{\mathbf{k}}$, with $\zeta_{\mathbf{k}} = \sqrt{(J_K/8)^2 + |\varepsilon_{\mathbf{k}}|^2}$. Here upper ($+$) and lower ($-$) bands correspond to the band index $m=1$ and $m=3$ in the previous trigonometric solutions, with the respective spectral weights  $|u_{\pm}|^2 = \frac{1}{2} \left( 1 \pm \frac{J_K}{8\zeta_{\mathbf{k}}} \right)$.
While the bare spinon band $E_f^{(0)} = \lambda_f$ is pinned at the Fermi level $E_F=0$ at exact half-filling to satisfy the single-occupation constraint, corresponding to the band index $m=2$.

Now we focus on the critical point $J_c$ by considering  the limit $r\rightarrow 0^{+}$ from the Kondo phase ($J_K>J_c$).
The second-order perturbation results in the following three hybridizing bands:
\begin{align}
    E_{c\pm}(r) &\approx E_{c\pm}^{(0)} + r^2 \frac{J_K^2 |u_{\pm}|^2}{E_{c\pm}^{(0)} - \lambda_f} + {\cal O}(r^3), \\
    E_f(r) &\approx \lambda_f + r^2 J_K^2 \left[ \frac{|u_+|^2}{\lambda_f - E_{c+}^{(0)}} + \frac{|u_-|^2}{\lambda_f - E_{c-}^{(0)}} \right]+ {\cal O}(r^3) .
\end{align}
Notice again that the single-occupancy constraint $\lim_{r\to 0^+} \sum_m n_F(E_m) \frac{\partial E_m}{\partial \lambda_f} = L$ dictates $n_F(\lambda_f) = 1/2$, enforcing $\lambda_f = 0$ at zero temperature, and the half-filling condition $\lim_{r\to 0^+} \sum_m n_F(E_m)(-1) \frac{\partial E_m}{\partial \mu} = 2L$ leads to $n_F(E_{c+}^{(0)}) + n_F(E_{c-}^{(0)}) = 1$ or $\mu = J_c/8$.
Then the unperturbed bands collapse to $E_{c\pm}^{(0)} = \pm \zeta_{\mathbf{k}}$, and
 \begin{equation}
   \lim_{r\to 0^+}\lim_{T\to 0} \sum_m n_F(E_m) \frac{1}{r} \frac{\partial E_m}{\partial r} = 0 \cdot \left( \frac{J_c^2}{\zeta_{\mathbf{k}}}+ \frac{J_c^3}{8\zeta_{\mathbf{k}}^2} \right) + \frac{1}{2} \cdot \left( -\frac{J_c^3}{4\zeta_{\mathbf{k}}^2} \right) + 1 \cdot \left( -\frac{J_c^2}{\zeta_{\mathbf{k}}} + \frac{J_c^3}{8\zeta_{\mathbf{k}}^2} \right) = -\frac{J_c^2}{\zeta_{\mathbf{k}}}.
\end{equation}
Substituting this into Eq.~(S8) ( with $Z=1$ and $\sum_\sigma = 2$), we obtain
\begin{equation}
    \frac{2}{J_c} = \frac{1}{L} \sum_{\mathbf{k}} \frac{1}{\sqrt{(J_c/8)^2 + |\varepsilon_{\mathbf{k}}|^2}}.
\end{equation}

In the continuous limit ($L \to \infty$), the $\mathbf k$-summation is transformed to the integration over energy $\varepsilon$:
\begin{equation}
    \frac{2}{J_c} = \int_0^{3t} \frac{\rho(\varepsilon)}{\sqrt{\varepsilon^2 + (J_c/8)^2}} d\varepsilon.
\end{equation}
In the above,
\begin{eqnarray}
\rho(\varepsilon) = \frac{2\varepsilon}{\pi^2 t^2 \sqrt{Z_0(\varepsilon)}} K\left(\sqrt{\frac{Z_1(\varepsilon)}{Z_0(\varepsilon)}}\right),
\end{eqnarray}
where $K(x)$ is the complete elliptic integral of the first kind. The piecewise functions $Z_0, Z_1$ are defined as\cite{SMNetoRMP2009}:
For $0 \leq \varepsilon \leq t$: $Z_0(\varepsilon) = \left(1 + \frac{\varepsilon}{t}\right)^2 - \frac{1}{4}\left[\left(\frac{\varepsilon}{t}\right)^2 - 1\right]^2$, $Z_1(\varepsilon) = 4\frac{\varepsilon}{t}$.
For $t < \varepsilon \leq 3t$: $Z_0(\varepsilon) = 4\frac{\varepsilon}{t}$, $Z_1(\varepsilon) = \left(1 + \frac{\varepsilon}{t}\right)^2 - \frac{1}{4}\left[\left(\frac{\varepsilon}{t}\right)^2 - 1\right]^2$.
Solving Eq.~(S13) yields $J_c \approx 2.564 t$. This value of $J_c$ is in perfect agreement with the low-temperature numerical result $J_c=2.575 t$ reported in the main text.

\subsection*{2. \textbf{B}-configuration}

In contrast to the \textbf{T}-configuration, the \textbf{B}-configuration preserves the inversion symmetry and the local potential is uniform $\Delta E_A = \Delta E_B = J_K/4$. The unperturbed conduction bands are $E_{c\pm}^{(0)} = -\mu + \frac{J_K}{4} \pm |\varepsilon_{\mathbf{k}}|$ with equal spectral weights $|u_{\pm}|^2 = 1/2$, implying that the itinerant electrons in the decoupled phase remain in the gapless Dirac liquid at half-filling.

In the Kondo phase with a finite $r$, the Kondo coupling $J_K$ induces the momentum-dependent hybridization $J_K r e^{\pm i k_x a_0/2}$ between the itinerant electrons and $f$-spinons. Similarly, the half-filling condition and the single-occupancy constraints lead to $\mu = J_c/4$ and $\lambda_f = 0$.  Near the critical point $J_c$, the three hybridizing bands are expanded up to the second order of $r$ as
\begin{align}
    E_{c+}(r) &\approx |\varepsilon_{\mathbf{k}}| + r^2 J_c^2 \left[ \frac{1}{|\varepsilon_{\mathbf{k}}|} - t \frac{\text{Re}(e^{-i k_x a_0} \gamma_{\mathbf{k}})}{|\varepsilon_{\mathbf{k}}|^2} \right] + {\cal O}(r^3), \\
    E_{c-}(r) &\approx -|\varepsilon_{\mathbf{k}}| - r^2 J_c^2 \left[ \frac{1}{|\varepsilon_{\mathbf{k}}|} + t \frac{\text{Re}(e^{-i k_x a_0} \gamma_{\mathbf{k}})}{|\varepsilon_{\mathbf{k}}|^2} \right] + {\cal O}(r^3), \\
    E_f(r) &\approx 0 + r^2 2J_c^2 \left[ t \frac{\text{Re}(e^{-i k_x a_0} \gamma_{\mathbf{k}})}{|\varepsilon_{\mathbf{k}}|^2} \right] + {\cal O}(r^3).
\end{align}
Following the same strategy described in the \textbf{T}-configuration, we obtain the following equation for $J_c$ in the \textbf{B}-configuration:
\begin{equation}
    \frac{2}{J_c} = \frac{1}{L} \sum_{\mathbf{k}} \left( \frac{1}{|\varepsilon_{\mathbf{k}}|} + t \frac{|\text{Re}(e^{-i k_x a_0} \gamma_{\mathbf{k}})|}{|\varepsilon_{\mathbf{k}}|^2} \right).
\end{equation}
Using $|\gamma_{\mathbf{k}}|^2 = |\varepsilon_{\mathbf{k}}|^2/t^2$ and taking the continuous limit $L\rightarrow \infty$, we have
\begin{equation}
    \frac{2}{J_c} = \int_0^{3t} \frac{\rho(\varepsilon)}{\varepsilon} d\varepsilon \times \left(1 + \frac{2}{\pi}\right),
\end{equation}
with $\rho(\varepsilon)$ introduced previously. Numerical integration yields $\int_0^{3t} \frac{\rho(\varepsilon)}{\varepsilon} d\varepsilon \approx \frac{0.730}{t}$, giving rise to $J_c \approx 1.674 t$.

\subsection*{3. \textbf{H}-configuration}

Similar to the \textbf{B}-configuration, the \textbf{H}-configuration preserves inversion symmetry and the local potential is uniform $\Delta E_A = \Delta E_B = 3J_K/4$. At the critical point the half-filling condition and the single-occupancy constraint lock $\mu = 3J_c/4$ and $\lambda_f = 0$. Consequently, the unperturbed Dirac fermions are gapless with $E_{c\pm}^{(0)} \to \pm |\varepsilon_{\mathbf{k}}|$ and $|u_{\pm}|^2 = 1/2$.

Approaching the critical point $J_c$ from the Kondo phase, the three hybridizing bands are expanded up to the second order of $r$ as
\begin{align}
    E_{c+}(r) &\approx |\varepsilon_{\mathbf{k}}| + r^2 J_c^2 \left[ \frac{|\varepsilon_{\mathbf{k}}|}{t^2} - t \frac{\text{Re}(\gamma_{\mathbf{k}}^3)}{|\varepsilon_{\mathbf{k}}|^2} \right] + {\cal O}(r^3), \\
    E_{c-}(r) &\approx -|\varepsilon_{\mathbf{k}}| - r^2 J_c^2 \left[ \frac{|\varepsilon_{\mathbf{k}}|}{t^2} + t \frac{\text{Re}(\gamma_{\mathbf{k}}^3)}{|\varepsilon_{\mathbf{k}}|^2} \right]+ {\cal O}(r^3), \\
    E_f(r) &\approx 0 + r^2 2J_c^2 \left[ t \frac{\text{Re}(\gamma_{\mathbf{k}}^3)}{|\varepsilon_{\mathbf{k}}|^2} \right]+ {\cal O}(r^3).
\end{align}
The corresponding $J_c$ for the \textbf{H}-configuration  is then determined by the following equation
\begin{equation}
    \frac{2}{J_c} = \frac{1}{3L} \sum_{\mathbf{k}} \left( \frac{|\varepsilon_{\mathbf{k}}|}{t^2} + t \frac{|\text{Re}(\gamma_{\mathbf{k}}^3)|}{|\varepsilon_{\mathbf{k}}|^2} \right).
\end{equation}
Its  continuous version is
\begin{eqnarray}
\frac{2}{J_c} = \frac{1}{3t^2} \int_0^{3t} \rho(\varepsilon) \varepsilon d\varepsilon \times \left(1 + \frac{2}{\pi}\right).
\end{eqnarray}
This result leads to $J_c \approx 2.494 t$ in the \textbf{H}-configuration.

\section*{SM-D: Absence of Kondo gap due to band crossings}
\label{sm:band_details}

 In Kondo systems, band repulsion, rather than band crossing, is visually ascribed to the hybridization nature of Kondo coupling ensuring the existence of a Kondo gap. In the main text, however, the numerical results of the hybridized quasiparticle band structures for the \textbf{H}- and \textbf{B}-configurations exhibit distinct metallic behavior associated with band crossings, leading to the absence of Kondo gap. This peculiar feature is attributed to complete Kondo decoupling of a conduction band along some specific momentum paths ( marked by the ``$-$'' sign) although the whole system is still in the Kondo phase. The appearance of the decoupled band allows band crossings without repulsion with the remaining hybridizing band along these momentum paths. Here we explain how this scenario takes place in the present HKLM due to the inversion or mirror symmetry.

\subsection*{1. Antibonding decoupling in the H-Configuration}

The effective mean-field Hamiltonian matrix for the \textbf{H}-configuration in the original sublattice basis $(c_{A\mathbf{k}}, c_{B\mathbf{k}}, f_{\mathbf{k}})^T$ reads
\begin{equation}
H_{\mathbf{H}}(\mathbf{k}) = \begin{pmatrix}
-\mu+\frac{3}{4}J_K & \varepsilon_{c,\mathbf{k}} & -J_K r \gamma_{\mathbf{k}}^* \\
\varepsilon_{c,\mathbf{k}}^* & -\mu+\frac{3}{4}J_K & -J_K r \gamma_{\mathbf{k}} \\
-J_K r \gamma_{\mathbf{k}} & -J_K r \gamma_{\mathbf{k}}^* & \lambda_f
\end{pmatrix},
\end{equation}
with $\varepsilon_{c,\mathbf{k}} = -t \gamma_{\mathbf{k}}$. This Hamiltonian is $C_6$ invariant preserving the inversion symmetry under $c_{A\mathbf {k}}\leftrightarrow c_{B\mathbf{k}}$.
We denote $\gamma_{\mathbf{k}} = |\gamma_{\mathbf{k}}|e^{i\theta_{\mathbf{k}}}$ and introduce a unitary transformation $U$ which sends $(c_{A\mathbf {k}}, c_{B\mathbf{k}})$ to
\begin{eqnarray}
 c_{\pm,\mathbf{k}} = \frac{1}{\sqrt{2}} \left( e^{i\theta_{\mathbf{k}}} c_{A\mathbf{k}} \pm e^{-i\theta_{\mathbf{k}}} c_{B\mathbf{k}} \right).
\end{eqnarray}
Here $c_{\pm,\mathbf {k}}$ describe the bonding/antibonding superposition states although they are not necessarily the diagonal basis of the Hamiltonian.  The transformed Hamiltonian matrix $\tilde{H}_{\mathbf{H}} = U^\dagger H_{\mathbf{H}} U$ is given by
\begin{eqnarray}
\tilde{H}_{\mathbf{H}} = \begin{pmatrix}
\varepsilon_c - t|\gamma_{\mathbf{k}}|\cos(3\theta_{\mathbf{k}}) & it|\gamma_{\mathbf{k}}|\sin(3\theta_{\mathbf{k}}) & -\sqrt{2}J_K r|\gamma_{\mathbf{k}}| \\
-it|\gamma_{\mathbf{k}}|\sin(3\theta_{\mathbf{k}}) & \varepsilon_c + t|\gamma_{\mathbf{k}}|\cos(3\theta_{\mathbf{k}}) & 0 \\
-\sqrt{2}J_K r|\gamma_{\mathbf{k}}| & 0 & \lambda_f
\end{pmatrix},
\label{eq:H_new}
\end{eqnarray}
with $\varepsilon_c = -\mu + \frac{3}{4}J_K$. Then, the direct coupling between the $f$-spinon and the antibonding state ($c_{-,\mathbf k}$) disappears. Of course, the kinetic hopping between the bonding-antibonding states persists, but becomes momentum-dependent  proportional to $|\gamma_{\mathbf{k}}|\sin(3\theta_{\mathbf{k}})$. Therefore, the strict decoupling of the antibonding band from the $f$-spinon occurs when $|\gamma_{\mathbf{k}}|\sin(3\theta_{\mathbf{k}}) = 0$. Here, three situations should be distinguished:
\begin{itemize}
    \item Essential degeneracy at $K$: At the Dirac node where $|\gamma_{\mathbf{K}}|=0$, both the bare dispersion and the Kondo hybridization term $-\sqrt{2}J_K r|\gamma_{\mathbf{K}}|$ in Eq.~(\ref{eq:H_new}) vanish. The lattice destructive interference totally nullifies the Kondo interaction, inherently preserving the degeneracy protected by the crystalline $C_{6v}$ symmetry.
    \item Decoupling along $M \rightarrow K$: Along this boundary path, $\theta_{\mathbf{k}} = -\pi/3$, yielding $\cos(3\theta_{\mathbf{k}}) = -1$ and $\sin(3\theta_{\mathbf{k}}) = 0$. The antibonding state becomes an eigenstate which completely decouples from the remaining ones and evolves with a pure kinetic dispersion $E_- = \varepsilon_c - t|\gamma_{\mathbf{k}}|$. It remains at low energy without intersecting other bands (corresponding to the isolated solid line marked with ``$-$'' at the bottom of the band spectra).
    \item Band crossing on $K \rightarrow \Gamma$: On this path ($k_x = 0$), $\theta_{\mathbf{k}} = 0$, yielding $\cos(3\theta_{\mathbf{k}}) = 1$ and $\sin(3\theta_{\mathbf{k}}) = 0$. The decoupled antibonding state possesses a band dispersion $E_- = \varepsilon_c + t|\gamma_{\mathbf{k}}|$ which starts from the Dirac $K$ point and climbs steeply toward $\Gamma$. It degenerates with the lower hybridizing band at the $K$ point showing a Dirac crossing and inevitably penetrates the upper heavy fermion band, the latter is pushed upward by the Kondo repulsion, yielding the band crossing along $K \rightarrow \Gamma$ due to the interplay of symmetry and Kondo coupling.
\end{itemize}

\subsection*{2. Odd-parity decoupling in the \textbf{B}-configuration}

For the \textbf{B}-configuration, the $C_{3v}$ rotational symmetry is broken, while the Kondo coupling shows an inversion symmetry, or specifically, the ${\cal M}_y$ mirror symmetry determined by the local potential $\Delta E_{A/B}=J_K/4 $ and the hybridization form factor ${\cal W}_{A/B}= e^{i \pm k_x / 2}$. This symmetry allows the definition of the even(+)/odd(-) parity states using the basis $c_{\pm,\mathbf{k}} = \frac{1}{\sqrt{2}} \left( e^{-ik_x/2} c_{A,\mathbf{k}} \pm e^{ik_x/2} c_{B,\mathbf{k}} \right)$. The transformed Hamiltonian matrix in this basis reads
\begin{equation}
\tilde{H}_{\mathbf{B}} = \begin{pmatrix}
\varepsilon_c' + \eta_{\mathbf{k}} & i\xi_{\mathbf{k}} & -\sqrt{2}J_K r \\
-i\xi_{\mathbf{k}} & \varepsilon_c' - \eta_{\mathbf{k}} & 0 \\
-\sqrt{2}J_K r & 0 & \lambda_f
\end{pmatrix},
\end{equation}
where, $\varepsilon_c' = -\mu + \frac{1}{4}J_K$, $\eta_{\mathbf{k}} = \mathrm{Re}(\varepsilon_{c,\mathbf{k}}e^{-ik_x}) = -t [1+2\cos(1.5k_x)\cos(\frac{\sqrt{3}}{2}k_y)]$, and $\xi_{\mathbf{k}} = -\mathrm{Im}(\varepsilon_{c,\mathbf{k}}e^{-ik_x}) = -2t\sin(\frac{3}{2}k_x)\cos(\frac{\sqrt{3}}{2}k_y)$. Therefore, the effective direct Kondo coupling is momentum-independent for the even-parity state and absent for the odd-parity state. Accordingly, three situations are distinguished:

\begin{itemize}
    \item Gap opening at $K$: At the $K$ point, $\eta_{K} = \xi_{K} = 0$. The odd-parity state decouples from both the even-parity state and the $f$-spinon. The momentum-independent hybridization between the even-parity state and $f$-spinon lifts the original twofold degeneracy, opening a Kondo gap at $K$.
    \item Avoided crossing on $M \rightarrow K$: Since $\xi_{\mathbf{k}} \neq 0$ along this boundary path, the odd-parity band fails to decouple. All three bands strongly mix and repel each other without band crossings.
    \item Band crossings due to mirror symmetry: Despite the gap opening at $K$, the \textbf{B}-configuration retains its overall metallicity at half-filling by relocating the semimetallic band crossings. The condition for complete decoupling ($\xi_{\mathbf{k}} \equiv 0$) is revived on two specific paths: $K \rightarrow \Gamma$ ($k_x = 0$) and $M' \rightarrow K'$ ($k_x = 2\pi/3$), respectively. Along $K \rightarrow \Gamma$, the decoupled odd-parity state shows a band dispersion $E_- = \varepsilon_c' + t\vert{}\gamma_{\mathbf{k}}\vert{}$ which rapidly ascends to intersect the hybridized band, creating a band crossing. Along $M' \rightarrow K'$, the dispersion of the decoupled odd-parity state follows $E_- = \varepsilon_c' - t\vert{}\gamma_{\mathbf{k}}\vert{}$ which increases smoothly from the $M'$ valley and intersects the hybridized band that is pushed downward by the Kondo repulsion, resulting in a second band crossing.
\end{itemize}


\begin{thebibliography}{0}%
\makeatletter
\providecommand \@ifxundefined [1]{%
 \@ifx{#1\undefined}
}%
\providecommand \@ifnum [1]{%
 \ifnum #1\expandafter \@firstoftwo
 \else \expandafter \@secondoftwo
 \fi
}%
\providecommand \@ifx [1]{%
 \ifx #1\expandafter \@firstoftwo
 \else \expandafter \@secondoftwo
 \fi
}%
\providecommand \natexlab [1]{#1}%
\providecommand \enquote  [1]{``#1''}%
\providecommand \bibnamefont  [1]{#1}%
\providecommand \bibfnamefont [1]{#1}%
\providecommand \citenamefont [1]{#1}%
\providecommand \href@noop [0]{\@secondoftwo}%
\providecommand \href [0]{\begingroup \@sanitize@url \@href}%
\providecommand \@href[1]{\@@startlink{#1}\@@href}%
\providecommand \@@href[1]{\endgroup#1\@@endlink}%
\providecommand \@sanitize@url [0]{\catcode `\\12\catcode `\$12\catcode
  `\&12\catcode `\#12\catcode `\^12\catcode `\_12\catcode `\%12\relax}%
\providecommand \@@startlink[1]{}%
\providecommand \@@endlink[0]{}%
\providecommand \url  [0]{\begingroup\@sanitize@url \@url }%
\providecommand \@url [1]{\endgroup\@href {#1}{\urlprefix }}%
\providecommand \urlprefix  [0]{URL }%
\providecommand \Eprint [0]{\href }%
\providecommand \doibase [0]{https://doi.org/}%
\providecommand \selectlanguage [0]{\@gobble}%
\providecommand \bibinfo  [0]{\@secondoftwo}%
\providecommand \bibfield  [0]{\@secondoftwo}%
\providecommand \translation [1]{[#1]}%
\providecommand \BibitemOpen [0]{}%
\providecommand \bibitemStop [0]{}%
\providecommand \bibitemNoStop [0]{.\EOS\space}%
\providecommand \EOS [0]{\spacefactor3000\relax}%
\providecommand \BibitemShut  [1]{\csname bibitem#1\endcsname}%
\let\auto@bib@innerbib\@empty
\end{thebibliography}%


\begin{thebibliography}{99}

\bibitem{StewartRMP1984}
G. Stewart, Heavy-fermion systems, Rev. Mod. Phys. \textbf{56}, 755 (1984).

\bibitem{StewartRMP2001}
G. Stewart, Non-Fermi-liquid behavior in $d$- and $f$-electron metals, Rev. Mod. Phys. \textbf{73}, 797 (2001).

\bibitem{Colemanbook2007}
P. Coleman, Heavy fermions: Electrons at the edge of magnetism, in \textit{Handbook of Magnetism and Advanced Magnetic Materials}, edited by H. Kronm\"uller and S. Parkin (Wiley, New York, 2007).

\bibitem{VanoNature2021}
V. Va\v{n}o, M. Amini, S. C. Ganguli, G. Chen, J. L. Lado, S. Kezilebieke, and P. Liljeroth, Artificial heavy fermions in a van der Waals heterostructure, Nature \textbf{599}, 582 (2021).

\bibitem{ChenPRB2021}
C. Chen, I. Sodemann, and P. A. Lee, Competition of spinon Fermi surface and heavy Fermi liquid states from the periodic Anderson to the Hubbard model, Phys. Rev. B \textbf{103}, 085128 (2021).

\bibitem{RamireaPRL2021}
A. Ramires and J. L. Lado, Emulating heavy fermions in twisted trilayer graphene, Phys. Rev. Lett. \textbf{127}, 026401 (2021).

\bibitem{DalalPRR2021}
A. Dalal and J. Ruhman, Orbitally selective Mott phase in electron-doped twisted transition-metal dichalcogenides: a possible realization of the Kondo lattice model, Phys. Rev. Res. \textbf{3}, 043173 (2021).

\bibitem{KumarPRB2022}
A. Kumar, N. C. Hu, A. H. MacDonald, and A. C. Potter, Gate-tunable heavy fermion quantum criticality in a moir\'{e} Kondo lattice, Phys. Rev. B \textbf{106}, L041116 (2022).

\bibitem{ShiPRB2022}
H. Shi and X. Dai, Heavy fermion representation for twisted bilayer graphene systems, Phys. Rev. B \textbf{106}, 245129 (2022).


\bibitem{SongPRL2022}
Z.-D. Song and B. A. Bernevig, Magic-angle twisted bilayer graphene as a topological heavy fermion problem, Phys. Rev. Lett. \textbf{129}, 047601 (2022).

\bibitem{ZhaoNature2023}
W. Zhao, B. Shen, Z. Tao, Z. Han, K. Kang, K. Watanabe, T. Taniguchi, K. F. Mak, and J. Shan, Gate-tunable heavy fermions in a moir\'{e} Kondo lattice, Nature \textbf{616}, 61 (2023).


\bibitem{HuSA2023}
H. Hu and Q. Si, Coupled topological flat and wide bands: Quasiparticle formation and destruction, Sci. Adv. \textbf{9}, eadg0028 (2023).

\bibitem{GuerciSA2023}
D. Guerci, J. Wang, J. Zang, J. Cano, J. H. Pixley, and A. J. Millis, Chiral Kondo lattice in doped MoTe$_2$/WS$_2$ bilayers, Sci. Adv. \textbf{9}, eade7701 (2023).


\bibitem{PoseyNature2024}
V. A. Posey, S. Turkel, M. Rezaee, A. Devarakonda, \textit{et al.}, Two-dimensional heavy fermions in the van der Waals metal CeSiI, Nature \textbf{625}, 483 (2024).

\bibitem{FumegaNL2024}
A. O. Fumega and J. L. Lado, Nature of the unconventional heavy fermion Kondo state in monolayer CeSiI, Nano Lett. \textbf{24}, 4272 (2024).

\bibitem{ChenNC2024}
L. Chen, F. Xie, S. Sur, H. Hu, S. Paschen, J. Cano, and Q. Si, Emergent flat band and topological Kondo semimetal driven by orbital-selective correlations, Nat. Commun. \textbf{15}, 5242 (2024).

\bibitem{CheckelskyNRM2024}
J. G. Checkelsky, B. A. Bernevig, P. Coleman, Q. Si, and S. Paschen, Flat bands, strange metals and the Kondo effect, Nat. Rev. Mater. \textbf{9}, 509 (2024).

\bibitem{XiePRR2024}
F. Xie, L. Chen, and Q. Si, Kondo effect and its destruction in heterobilayer transition metal dichalcogenides, Phys. Rev. Res. \textbf{6}, 013219 (2024).

\bibitem{Mahankaliarxiv2025}
M. Mahankali, F. Xie, Y. Fang, L. Chen, \textit{et al.}, Correlated flat-band physics in a bilayer kagome metal based on compact molecular orbitals, arXiv:2503.09706 (2025).

\bibitem{XiaoNature2026}
J. Xiao, A. Inbar, J. Birkbeck, N. Gershon, Y. Zamir, Y. Vituri, T. Taniguchi, K. Watanabe, E. Berg, and S. Ilani, Imaging the flat bands of magic-angle graphene reshaped by interactions, Nature \textbf{653}, 68 (2026).

\bibitem{ZhongPRB2026}
H. Zhong, S. Yang, C. Cao, X.-Y. Feng, and J. Dai, Selective Kondo screening and strange metallicity in sliding Dirac semimetals, Phys. Rev. B \textbf{113}, 075111 (2026).

\bibitem{ColemanJPCM2001}
P. Coleman, C. P\'epin, Q. Si, and R. Ramazashvili, How do Fermi liquids get heavy and die?, J. Phys. Condens. Matter \textbf{13}, R723 (2001).

\bibitem{SiNature2001}
Q. Si, S. Rabello, K. Ingersent, and J. Smith, Locally critical quantum phase transitions in strongly correlated metals, Nature \textbf{413}, 804 (2001).

\bibitem{SiScience2010}
Q. Si and F. Steglich, Heavy fermions and quantum phase transitions, Science \textbf{329}, 1161 (2010).

\bibitem{LuttingerPR1960}
J. M. Luttinger, Fermi surface and some simple equilibrium properties of a system of interacting fermions, Phys. Rev. \textbf{119}, 1153 (1960).

\bibitem{LuttingerPR1960a}
J. M. Luttinger and J. C. Ward, Ground-state energy of a many-fermion system. II, Phys. Rev. \textbf{118}, 1417 (1960).

\bibitem{NakajimaPTP1960}
S. Nakajima, On the Fermi liquid theory of electrons in metals, Prog. Theor. Phys. \textbf{23}, 694 (1960).

\bibitem{MartinPRL1982}
R. M. Martin, Fermi-surface sum rule and its generalizations for interacting fermions, Phys. Rev. Lett. \textbf{48}, 362 (1982).

\bibitem{OshikawaPRL2000}
M. Oshikawa, Commensurability, excitation gap, and topology in quantum many-particle systems on a periodic lattice, Phys. Rev. Lett. \textbf{84}, 1535 (2000).

\bibitem{OshikawaPRL2000Kondo}
M. Oshikawa, Topological approach to Luttinger's theorem and the Fermi surface of a Kondo lattice, Phys. Rev. Lett. \textbf{84}, 3370 (2000).


\bibitem{HazraPRR2021}
T. Hazra and P. Coleman, Luttinger sum rules and spin fractionalization in the $SU(N)$ Kondo lattice, Phys. Rev. Res. \textbf{3}, 033284 (2021).


\bibitem{VojtaJLTP2010}
M. Vojta, Orbital-selective Mott transitions: Heavy fermions and beyond, J. Low Temp. Phys. \textbf{161}, 203 (2010).

\bibitem{DavePRL2013}
K. B. Dave, P. W. Phillips, and C. L. Kane, Absence of Luttinger's theorem due to zeros in the single-particle Green function, Phys. Rev. Lett. \textbf{110}, 090403 (2013).

\bibitem{BlasonPRB2023}
A. Blason and M. Fabrizio, Unified role of Green's function poles and zeros in correlated topological insulators, Phys. Rev. B \textbf{108}, 125115 (2023).

\bibitem{GavenskyPRL2023}
L. P. Gavensky, S. Sachdev, and N. Goldman, Connecting the many-body Chern number to Luttinger's theorem through St\v{r}eda's formula, Phys. Rev. Lett. \textbf{131}, 236601 (2023).


\bibitem{SenthilPRL2003}
T. Senthil, S. Sachdev, and M. Vojta, Fractionalized Fermi liquids, Phys. Rev. Lett. \textbf{90}, 216403 (2003).


\bibitem{SenthilPRB2004}
T. Senthil, M. Vojta, and S. Sachdev, Weak magnetism and non-Fermi liquids near heavy-fermion critical points, Phys. Rev. B \textbf{69}, 035111 (2004).


\bibitem{BalentsNature2010}
L. Balents, Spin liquids in frustrated magnets, Nature \textbf{464}, 199 (2010).

\bibitem{ZhouRMP2017}
Y. Zhou, K. Kanoda, and T.-K. Ng, Quantum spin liquid states, Rev. Mod. Phys. \textbf{89}, 025003 (2017).

\bibitem{KivelsonNRP2023}
S. A. Kivelson and S. L. Sondhi, 50 years of quantum spin liquids, Nat. Rev. Phys. \textbf{5}, 368 (2023).

\bibitem{BonettiRPP2026}
P. M. Bonetti, M. Christos, A. Nikolaenko, A. A. Patel, and S. Sachdev, Fractionalized Fermi liquids and the cuprate phase diagram, Rep. Prog. Phys. \textbf{89}, 044501 (2026).

\bibitem{ZhangPRR2020}
Y.-H. Zhang and S. Sachdev, From the pseudogap metal to the Fermi liquid using ancilla qubits, Phys. Rev. Res. \textbf{2}, 023172 (2020).

\bibitem{ZhangPRB2020}
Y.-H. Zhang and S. Sachdev, Deconfined criticality and ghost Fermi surfaces at the onset of antiferromagnetism in a metal, Phys. Rev. B \textbf{102}, 155124 (2020).

\bibitem{ColemanPRL2022}
P. Coleman, A. Panigrahi, and A. Tsvelik, Solvable 3D Kondo lattice exhibiting pair density wave, odd-frequency pairing, and order fractionalization, Phys. Rev. Lett. \textbf{129}, 177601 (2022).

\bibitem{DanuPRL2020}
B. Danu, M. Vojta, F. F. Assaad, and T. Grover, Kondo breakdown in a spin-$1/2$ chain of adatoms on a Dirac semimetal, Phys. Rev. Lett. \textbf{125}, 206602 (2020).


\bibitem{ManulPRB1999}
L. O. Manuel and H. A. Ceccatto, Magnetic and quantum disordered phases in triangular-lattice Heisenberg antiferromagnets, Phys. Rev. B \textbf{60}, 9489 (1999).

\bibitem{ZhuPRB2015}
Z. Zhu and S. R. White, Spin liquid phase of the $S=\frac{1}{2}$ $J_1-J_2$ Heisenberg model on the triangular lattice, Phys. Rev. B \textbf{92}, 041105(R) (2015).

\bibitem{IqbalPRB2016}
Y. Iqbal, W.-J. Hu, R. Thomale, D. Poilblanc, and F. Becca, Spin liquid nature in the Heisenberg $J_1$-$J_2$ triangular antiferromagnet, Phys. Rev. B \textbf{93}, 144411 (2016).

\bibitem{HuPRL2019}
S. Hu, W. Zhu, S. Eggert, and Y.-C. He, Dirac spin liquid on the spin-$1/2$ triangular Heisenberg antiferromagnet, Phys. Rev. Lett. \textbf{123}, 207203 (2019).

\bibitem{SM}
See Supplemental Material at [URL will be inserted by publisher] for details on the numerical influence of $J'_H$, the eigenvalues of the Hamiltonian matrix, the critical Kondo coupling, as well as the origin of band crossings.

\bibitem{AbrikosovPRF1965}
A. A. Abrikosov, Electron scattering on magnetic impurities in metals and anomalous resistivity effects, Phys. Phys. Fiz. \textbf{2}, 5 (1965).

\bibitem{LeeRMP2006}
P. A. Lee, N. Nagaosa, and X.-G. Wen, Doping a Mott insulator: Physics of high-temperature superconductivity, Rev. Mod. Phys. \textbf{78}, 17 (2006).

\bibitem{FradkinPRD1979}
E. Fradkin and S. H. Shenker, Phase diagrams of lattice gauge theories with Higgs fields, Phys. Rev. D \textbf{19}, 3682 (1979).

\bibitem{Colemanbook2016}
P. Coleman, \textit{Introduction to Many-Body Physics} (Cambridge University Press, Cambridge, England, 2015).

\bibitem{RaczkowskiPRB2022}
M. Raczkowski, B. Danu, and F. F. Assaad, Breakdown of heavy quasiparticles in a honeycomb Kondo lattice: A quantum Monte Carlo study, Phys. Rev. B \textbf{106}, L161115 (2022).

\bibitem{note2}
In the magnetic ordered metallic phase the Kondo coupling is marginal thereby opening the possibility of Kondo delocalization in the weak Kondo-coupling limit\cite{YamamotoPRL2007}.

\bibitem{YamamotoPRL2007}
S. J. Yamamoto and Q. Si, Fermi surface and antiferromagnetism in the Kondo lattice: an asymptotically exact solution in $d>1$ dimensions, Phys. Rev. Lett. \textbf{99}, 016401 (2007).

\bibitem{SemenoffPRL1984}
G. W. Semenoff, Condensed-matter simulation of a three-dimensional anomaly, Phys. Rev. Lett. \textbf{53}, 2449 (1984).

\bibitem{NetoRMP2009}
A. H. Castro Neto, F. Guinea, N. M. R. Peres, K. S. Novoselov, and A. K. Geim, The electronic properties of graphene, Rev. Mod. Phys. \textbf{81}, 109 (2009).

\bibitem{WithoffPRL1990}
D. Withoff and E. Fradkin, Phase transitions in gapless Fermi systems with magnetic impurities, Phys. Rev. Lett. \textbf{64}, 1835 (1990).

\bibitem{GonzalezPRB1998}
C. Gonzalez-Buxton and K. Ingersent, Renormalization-group study of Anderson and Kondo impurities in gapless Fermi systems, Phys. Rev. B \textbf{57}, 14254 (1998).

\bibitem{NozieresJP1980}
P. Nozi\`eres and A. Blandin, Kondo effect in real metals, J. Phys. (Paris) \textbf{41}, 193 (1980).

\bibitem{AndreiPRL1984}
N. Andrei and C. Destri, Solution of the multichannel Kondo problem, Phys. Rev. Lett. \textbf{52}, 364 (1984).

\bibitem{TsvelickJSP1985}
A. M. Tsvelick and P. B. Wiegmann, Exact solution of the multi-channel Kondo problem, scaling, and integrability, J. Stat. Phys. \textbf{38}, 125 (1985).

\bibitem{Fradkin2010}
E. Fradkin, S. A. Kivelson, M. J. Lawler, J. P. Eisenstein, and A. P. Mackenzie, Nematic Fermi fluids in condensed matter physics, Annu. Rev. Condens. Matter Phys. \textbf{1}, 153 (2010).

\bibitem{PhillipsScience2022}
P. W. Phillips, N. E. Hussey, and P. Abbamonte, Stranger than metals, Science \textbf{377}, eabh4273 (2022).

\bibitem{HuNP2024}
H. Hu, L. Chen, and Q. Si, Quantum critical metals and loss of quasiparticles, Nat. Phys. \textbf{20}, 1863 (2024).

\bibitem{SouzaNP2026}
J. C. Souza, M. Haim, A. Gupta, M. Mahankali, \textit{et al.}, Origin of strange metallicity in a $d$-orbital kagome metal, Nat. Phys. \textbf{22}, 541 (2026).

\bibitem{MazzaNP2026}
F. Mazza, S. Biswas, X. Yan, A. Prokofiev, P. Steffens, Q. Si, F. F. Assaad, and S. Paschen, Quantum Fisher information in a strange metal, Nat. Phys. \textbf{22}, 1064 (2026).


\bibitem{YeNP2024}
L. Ye, S. Fang, M. Kang, J. Kaufmann, \textit{et al.}, Hopping frustration-induced flat band and strange metallicity in a kagome metal, Nat. Phys. \textbf{20}, 610 (2024).

\bibitem{NuckollsNature2026}
K. P. Nuckolls, N. Paul, A. Chen, F. Gaggioli, \textit{et al.}, Higher-dimensional Fermiology in bulk moir\'e metals, Nature \textbf{651}, 333 (2026).


\bibitem{ZhangPRL2026}
R.-H. Zhang, J.-H. Pan, Y.-N. Ren, C.-Y. Hao, \textit{et al.}, Kagomelike bands in graphene/$\text{WSe}_2$ heterostructure realized by strong interlayer hybridization, Phys. Rev. Lett. \textbf{137}, 096201 (2026).


\end{thebibliography}

\begin{thebibliography}{99}
\bibitem{SMManulPRB1999}
L. O. Manuel and H. A. Ceccatto, Magnetic and spin-liquid phases in the frustrated $t-J$ model on the triangular lattice, Phys. Rev. B \textbf{60}, 9489 (1999).

\bibitem{SMZhuPRB2015}
Z. Zhu and S. R. White, Spin liquid phase of the $S=1/2$ $J_1-J_2$ Heisenberg model on the triangular lattice, Phys. Rev. B \textbf{92}, 041105 (2015).

\bibitem{SMIqbalPRB2016}
Y. Iqbal, W.-J. Hu, R. Thomale, D. Poilblanc, and F. Becca, Spin liquid nature in the Heisenberg $J_1-J_2$ triangular antiferromagnet, Phys. Rev. B \textbf{93}, 144411 (2016).

\bibitem{SMHuPRL2019}
W.-J. Hu, S.-S. Gong, W. Zhu, and D. N. Sheng, Discovering spin-liquid fractional excitations in the frustrated Heisenberg model on the triangular lattice, Phys. Rev. Lett. \textbf{122}, 147201 (2019).

\bibitem{SMCardano2007}
G. Cardano, T. R. Witmer, and O. Ore, \textit{The rules of algebra: Ars Magna} (Courier Corporation, 2007).

\bibitem{SMLambert1906}
W. D. Lambert, A generalized trigonometric solution of the cubic equation, Am. Math. Mon. \textbf{13}, 73 (1906).

\bibitem{SMNetoRMP2009}
A. H. Castro Neto, F. Guinea, N. M. R. Peres, K. S. Novoselov, and A. K. Geim, The electronic properties of graphene, Rev. Mod. Phys. \textbf{81}, 109 (2009).

\end{thebibliography}
\end{document}